\documentclass{elsart}
\usepackage{graphicx}
\begin{document}

\begin{frontmatter}
\title{Electromagnetic corrections to the hadronic phase shifts
in low energy $\pi^+p$ elastic scattering}

\author[ZH]{A. Gashi}
\author[ZH]{E. Matsinos$^*$}
\author[AA]{G.C. Oades}
\author[ZH]{G. Rasche}
\author[CA]{W.S. Woolcock}

\address[ZH]{Institut f\"{u}r Theoretische Physik der
Universit\"{a}t,
Winterthurerstrasse 190, CH-8057 Z\"{u}rich, Switzerland}
\address[AA]{Institute of Physics and Astronomy, Aarhus University,
DK-8000 Aarhus C, Denmark}
\address[CA]{Department of Theoretical Physics, IAS,
The Australian National University, Canberra, ACT 0200, Australia}

\begin{abstract}

We calculate for the $s$-, $p_{1/2}$- and $p_{3/2}$-waves the electromagnetic 
corrections which must be subtracted from the nuclear phase shifts obtained
from the analysis of low energy $\pi^+p$ elastic scattering data, in order to
obtain hadronic phase shifts. The calculation uses relativised Schr\"{o}dinger
equations containing the sum of an electromagnetic potential and an effective
hadronic potential. We compare our results with those of previous calculations
and estimate the uncertainties in the corrections.\\
\noindent{\it PACS:} 13.75.Gx,25.80.Dj
\end{abstract}

\begin{keyword}$\pi N$ elastic scattering, $\pi N$ electromagnetic corrections,
$\pi N$ phase shifts
\end{keyword}

$^*$\address{ Present address: The KEY Institute for Brain-Mind Research, 
University Hospital of Psychiatry, Lenggstrasse 31, CH-8029 Z\"{u}rich,
Switzerland}

\end{frontmatter}

\section{Introduction}
 
The aim of this paper is to present a careful recalculation of the electromagnetic
corrections which have to be applied in the analysis of low energy $\pi^+p$
elastic scattering data in order to recover the hadronic phase shifts.

For these electromagnetic effects in $\pi N$ scattering ($\pi^+p$ elastic and
$\pi^-p$ elastic and charge exchange) two different approaches were developed:
the dispersion theory method and a method using potentials for the hadronic and
electromagnetic interactions in a relativised Schr\"{o}dinger equation (RSE for
short). By this we mean an extension of the Schr\"{o}dinger equation to the
relativistic domain. 

The dispersion theory approach was initiated by Sauter \cite{1,2}. It was then
extended by the NORDITA group. The final results for their corrections are given
in Tromborg {\it et al.} \cite{3}. In Ref.\cite{4} the same authors present a
detailed exposition of the underlying ideas and point out their limitations.

The potential theory approach was initiated by Oades and Rasche \cite{5,6} and
applied in detail by Zimmermann \cite{7,8} to an analysis of the then existing
data. His final results appear in the Landolt-B\"{o}rnstein tables on $\pi N$
scattering \cite{9}.

Both methods have their shortcomings. In the dispersion theory approach the
only contribution from $t$- and $u$-channel exchange which is taken into account
is that from $u$-channel nucleon exchange. The rest of the unphysical
contributions are omitted. Ref.\cite{4} states: `We are not able to determine or
even estimate this term, which is of course a serious drawback of our method'.
For this reason, Ref.\cite{4} makes it clear that the results of an analysis of
data for the three measured reactions which uses the corrections given there
cannot be used to draw any conclusion about isospin symmetry violation. It seems
that the fact that a dispersion theory method cannot give absolute values of the
electromagnetic corrections has not been sufficiently recognised. This is an
unsatisfactory situation, particularly for the analysis of the large amount of
data at low energies.

In the potential theory approach  of Refs.\cite{7,8} hadronic potentials were
constructed via RSEs to reproduce the hadronic phase shifts. These potentials
then were used in RSEs, containing also the electromagnetic potential, to
calculate the electromagnetic corrections. The idea is that by this addition of
an effective short range hadronic potential and the long range electromagnetic
potential we can estimate the corrections reliably. The weak point here is the 
use of a hadronic potential in a RSE. But it should be noted that\\ 
(i) the hadronic potentials are used only to calculate the corrections, not to 
calculate the hadronic quantities themselves;\\
(ii)  the potential model is the only method that can give a well defined
separation between hadronic quantities and their electromagnetic corrections.

It was indeed possible (Ref.\cite{8}) to analyse the best experimental data of
that time in an isospin invariant way, using energy dependent hadronic
potentials. The main reasons for recalculating the electromagnetic corrections
in the potential model approach are the following.\\
(i) The results need to be extended to lower energies, where there is now a 
large amount of new data obtained at pion factories.\\
(ii) The slight effect which a change of the hadronic input has on the
electromagnetic corrections needs to be controlled. This has not been done by
the NORDITA group.\\ 
(iii) The effect of some fine details of the electromagnetic interaction, which
might become important at the present precision of the experimental data, needs
to be checked. 
This also is not contained in the NORDITA results.\\
(iv) Instead of the energy dependent potentials used in Refs.\cite{7,8} we want
to use energy independent hadronic potentials to calculate the electromagnetic
corrections for pion laboratory kinetic energy $T_\pi \leq 100$ MeV.

In this paper we treat only the single channel problem of $\pi^+p$ elastic
scattering. It enables us to describe the formalism carefully and to
provide a survey of the relative importance of the contributions to the
corrections. In the next paper we extend the formalism to the two-channel
($\pi^-p, \pi^0n$) case and give the results for the electromagnetic corrections
which need to be applied in the analysis of low energy $\pi^-p$ elastic and
charge exchange scattering data.

In Section 2 we give a detailed description of our model for the electromagnetic
corrections to $\pi^+p$ scattering. In Section 3 we explain the method of
evaluation of the corrections in this model and in Section 4 we give the final
numerical results for these corrections. 

\section{Detailed description of the model}

For the purpose of calculating the electromagnetic corrections we describe low
energy $\pi^+p$ scattering by means of individual potentials for each of the
partial waves which are used in relativised Schr\"{o}dinger  equations (RSEs).
We therefore develop the model by motivating the construction of the RSEs 
and discussing the components of the potentials that appear in them.

The point charge Coulomb potential is 
\begin{equation}
V^{pc}(r)=\alpha /r , \label{eq:1}
\end{equation}
where $r$ is the distance between $\pi^+$ and $p$ in the c.m. system.
When the potential (\ref{eq:1}) is used in the nonrelativistic Schr\"{o}dinger
equation, one obtains the nonrelativistic point charge Coulomb amplitude
\begin{equation}
f_{NR}^{pc}=\frac{2\alpha m_c}{t}\exp{\{2i(\sigma_0)_{NR}-i\eta \ln{(\sin^2
{\frac{1}{2}}\theta)}\}} , \label{eq:2}
\end{equation}
where
\[
(\sigma_0)_{NR}=\arg{\Gamma(1+i\eta)} , 
\]
\begin{equation}
\eta=\frac{\alpha m_c}{q_c} , \label{eq:3}  
\end{equation}
\[
t=-2q_c^2(1-\cos{\theta}), 
\]
$\theta$ being  the c.m. scattering angle and $m_c$ the reduced mass of the
$\pi^+ p$ system:
\[
m_c=\frac{m_p\mu_c}{m_p+\mu_c}  ,
\]
with $m_p$, $\mu_c$ the masses of the proton and charged pion respectively.
The quantity $q_c$ is the c.m. momentum of the $\pi^+ p$ system. The Born
approximation to the point charge Coulomb amplitude is given by the expression
(\ref{eq:2}) without the phase factor.

To make the relativistic generalisation of these nonrelativistic results,
we start from the one photon exchange ($1\gamma E$) no-flip and spin-flip
amplitudes $f_{1\gamma E}$, $g_{1\gamma E}$, modified by the pion and proton
form factors $F^{\pi}$ and $F_1^p$,  $F_2^p$ respectively, and separate them
into three parts:
\begin{equation}
f_{1\gamma E}=f^{pc}_{1\gamma E}+f^{ext}_{1\gamma E}+f^{rel}_{1\gamma E} ,
\label{eq:4}
\end{equation}
\begin{equation}
g_{1\gamma E}=g^{pc}_{1\gamma E}+g^{ext}_{1\gamma E}+g^{rel}_{1\gamma E} ,
\label{eq:5}
\end{equation}
where
\begin{equation}
f^{pc}_{1\gamma E}=\frac{2\alpha m_c f_c}{t} \, , \,g^{pc}_{1\gamma E}=0 ,
\label{eq:6}
\end{equation}
\begin{equation}
f^{ext}_{1\gamma E}=\frac{2\alpha m_c f_c}{t} (F^{\pi}F_1^p-1) \, , \,
g^{ext}_{1\gamma E}=0 ,
\label{eq:7}
\end{equation}
\begin{equation}
f^{rel}_{1\gamma E}=\frac{\alpha}{2W}\{ \frac{W+m_p}{E+m_p}F_1^p+
2(W-m_p+\frac{t}{4(E+m_p)})F_2^p\}F^{\pi}  ,
\label{eq:8}
\end{equation}
\begin{equation}
g^{rel}_{1\gamma E}=\frac{i\alpha}{2W\tan{(\frac{1}{2}\theta})}\{ \frac{W+m_p}
{E+m_p}F_1^p+2(W+\frac{t}{4(E+m_p)})F_2^p\}F^{\pi}  
\label{eq:9}
\end{equation}
and 
\begin{equation} 
f_c=\frac{W^{2}-m_p^2-\mu_c^2}{2m_cW} \, , \, q_c^2=\frac{[W^2-(m_p-\mu_c)^2]
\,[W^2-(m_p+\mu_c)^2]}{4W^2} ,
\label{eq:10}
\end{equation}
\[
E=(m_p^2+q_c^2)^{1/2}  .
\]
The quantity $W$ is the total energy in the c.m. frame and the results in Eqs.
(\ref{eq:6})-(\ref{eq:10}) are fully relativistic.

We now note that the relativistic ${1\gamma E}$ point charge amplitude
$f^{pc}_{1\gamma E}$ in Eq.(\ref{eq:6}) is the Born approximation to
$f^{pc}_{NR}$ multiplied by the factor $f_c$. It is thus the Born approximation
to the amplitude obtained from the RSE
\begin{equation}
\{\, \bigtriangledown^2+q_c^2-2m_cf_{c}V^{pc}(r)\,\}\,\psi(\mathbf{r})=0 .
\label{eq:11}
\end{equation}
The inclusion of $f_c$ in Eq.(\ref{eq:11}) gives the unambiguous relativistic
generalisation of the nonrelativistic potential term $2m_cV^{pc}$. The factor
$f_c=1$ when $W=\mu_c+m_p$ and increases as $W$ increases. 
The RSE (\ref{eq:11}) is obtained in a more rigorous way by following the
standard route from the Bethe-Salpeter equation, making a three-dimensional
reduction, using only the leading part of the $1\gamma E$ diagram contribution
and converting to coordinate space. This is the method used in Ref.\cite{10};
the factor $f_c$ is just the quantity $a$ in Eqs.(2) and (4) of that paper. 

The full amplitude $f^{pc}_{REL}$ obtained from the RSE (\ref{eq:11}) is
\begin{equation}
f_{REL}^{pc}=\frac{2\alpha m_c f_c}{t}\exp{\{2i\sigma_0-i\eta f_c\ln{(\sin^2
{\frac{1}{2}}\theta)}\}} ,
\label{eq:12}
\end{equation}
where 
\begin{equation}
\sigma_l=\arg{\Gamma(l+1+i\eta f_c)} . \label{eq:13}
\end{equation}
The expression (\ref{eq:12}) is the point charge Coulomb amplitude to all orders
in $\eta$. It agrees with the results in Eqs.(\ref{eq:2}) and (\ref{eq:6}) in
the appropriate limits and generalises those results.

From the RSE (\ref{eq:11}) one obtains the radial RSEs for individual partial
waves:
 \begin{equation}
(\,\frac{d^2}{dr^2}-\frac{l(l+1)}{r^2}+q_c^2-2m_cf_{c}V^{pc}(r)\,)\,u_{l}(r)=0 .
\label{eq:14}
\end{equation}
The wavefunction regular at $r=0$ has the asymptotic behaviour as
$r\rightarrow\infty$ 
\[
\sin(q_cr-\eta f_c\ln (2q_cr)-l\pi/2+\sigma_l) ,
\]
where $\sigma_l$ is given by Eq.(\ref{eq:13}). We shall now proceed to add
other potentials to $V^{pc}$, so that the wavefunction regular at $r=0$ behaves
as $r\rightarrow\infty$ like 
\[
\sin(q_cr-\eta f_c\ln (2q_cr)-l\pi/2+\sigma_l+\delta_{l\pm}) .
\]
The notation $\delta_{l\pm}$ takes account of the fact that the extra phase
shifts, as well as the added potentials, in general depend on the orbital
angular momentum $l$ and on the total angular momentum $j=l\pm \frac{1}{2}$. 
We now take the sum over all partial waves to obtain the full no-flip and
spin-flip amplitudes $f$, $g$ respectively and use the identity
\[
e^{2i(\sigma+\delta)}-1=(e^{2i\sigma}-1)+e^{2i\sigma}(e^{2i\delta}-1) .
\]
Furthermore, as is customary practice, we remove the phase factor
$e^{2i\sigma_0}$ from both amplitudes; this does not affect any observable.
The result is 
\begin{equation}
f=f^{pc}+\sum_{l=0}^{\infty} e^{2i(\sigma_l-\sigma_0)}\{(l+1)f_{l+}+lf_{l-}\}P_l ,
\label{eq:15}
\end{equation}
\begin{equation}
g=i\sum_{l=1}^{\infty} e^{2i(\sigma_l-\sigma_0)}(f_{l+}-f_{l-})P_l^1 ,
\label{eq:16}
\end{equation}
with
\begin{equation}
f_{l\pm}=\frac{\exp {(2i\delta_{l\pm})}-1}{2iq_c} , \label{eq:17}
\end{equation}
\begin{equation}
f^{pc}=\frac{2\alpha m_c f_c}{t}\exp{\{-i\eta f_c\ln{(\sin^2{\frac{1}{2}}\theta)}\}} .
\label{eq:18}
\end{equation}
The potentials other than $V^{pc}$ are sufficiently well behaved as $r\rightarrow
\infty$  for the sums over partial waves to be convergent.

Two photon exchange contributions give a very small correction to the $1\gamma E$
amplitudes and can be neglected at the level of accuracy required for the present
calculations. However, even though vacuum polarisation is of higher order in
$\alpha$, it needs to be taken into account because it is of extremely long range 
compared with the hadronic interaction. For vacuum polarisation we used the
standard Uehling potential for point charges given by Durand \cite{11}: 
\begin{equation}
V^{vp}(r)=V^{pc}(r)I(2 m_e r) 2\alpha / 3\pi ,  \label{eq:19}
\end{equation}
where $m_e$ is the electron mass. The explicit integral representation of $I$
can be taken from Ref.\cite{11}. 
Since $V^{vp}$ is a potential of electromagnetic origin, we assume (as is done
in the treatment of $pp$ scattering) that, when $V^{vp}$ is added to $V^{pc}$
in the RSEs (14), the factor $f_c$ should multiply it as well. The tiny effect
of the extended charge distributions on  $V^{vp}$ can be neglected. 
The vacuum polarisation amplitude $f^{vp}$ is given with sufficient accuracy
for the analysis of current $\pi^+p$ data by the lowest order expression in
Eq.(12.2) of Ref.\cite{11}:
\begin{equation}
f^{vp}=-\frac{\alpha \eta f_c}{3\pi q_c}(1-\cos{\theta})^{-1}F(\cos{\theta}),
\label{eq:20}
\end{equation}
where
\[
F(\cos{\theta})=-\frac{5}{3}+X+(1-\frac{1}{2}X)(1+X)^{1/2}\ln{ \{ \frac{(1+X)
^{1/2}+1} {(1+X)^{1/2}-1}   \} },
\]
\[
X=-\frac{4m_e^2}{t}.
\]

The next step is to explicitly separate from the phase shifts $\delta_{l\pm}$
the very small phase shifts $\sigma_l^{ext}$, $\sigma_{l\pm}^{rel}$ and
$\sigma_l^{vp}$. The partial-wave projections of the amplitudes 
given in Eqs.(\ref{eq:7})-(\ref{eq:9}) and (\ref{eq:20}) lead to the results 
\begin{equation}
\sigma_l^{ext}=\alpha m_c f_c q_c\int_{-1}^{+1}dz \,t^{-1}P_{l}(z)(F_1^pF
^{\pi}-1) ,
\label{eq:21}
\end{equation}
\begin{eqnarray}
\sigma_{l\pm}^{rel}=-\frac{\alpha q_cm_p}{2W}\int_{-1}^{+1}dz \, P_{l}
(z)F_2^pF^{\pi} \pm \frac{\alpha q_c}{4W(l\pm 1/2+1/2)} \times \nonumber \\
\int_{-1}^{+1}dz(P_{l}^{'}(z)+P_{l\pm 1}^{'}(z))\{\frac{W+m_p}{E+m_p}F_1^p+
(W+\frac{t}{4(E+m_p)})2F_2^p\} F^{\pi} , \label{eq:22}
\end{eqnarray}
\begin{equation}
\sigma_{l}^{vp}=-\frac{\alpha \eta f_c}{3\pi}\int_{0}^{1}dy(1+\frac{1}{2}y)
(1-y)^{1/2}y^{-1}Q_l(1+ \nu y^{-1}) , \label{eq:23}
\end{equation}
where $\nu=2m_e^2/q_c^2$ and $Q_l$ is the Legendre function of the second kind. 
The expression in Eq.(\ref{eq:23}) comes from Eq.(8.2) of Ref.\cite{11}.
For $l=0,...,3$  this result was checked  by direct integration of the RSEs, 
integrating to 1000 fm because of the very long range of $V^{vp}$. The
expressions (\ref{eq:21})-(\ref{eq:23}) are completely sufficient for these
very small phase shifts.

In order to calculate the electromagnetic corrections (which we do for $l=0,1$)
it is necessary to construct potentials $V_{0,1}^{ext}$ and $V_{0+,1\pm}^{rel}$
which reproduce the phase shifts $\sigma_{0,1}^{ext}$ and $\sigma_{0+,1\pm}^{rel}$
with good accuracy up to $T_{\pi}=100$ MeV. The phase shifts themselves are
calculated from Eqs.(\ref{eq:21}) and (\ref{eq:22}) with the dipole form factors
\[
F_1^p(t)=(1-t/\Lambda_p^2)^{-2}  ,
\]
\[
F_2^p(t)=\frac{\kappa_p}{2m_p}F_1^p(t) ,
\]
\[
F^{\pi}(t)=(1-t/\Lambda_{\pi}^2)^{-2} ,
\]
where $\Lambda_p=805 \, \textnormal{MeV}  , \Lambda_{\pi}=1040 \, \textnormal{MeV}.$
The parameters $\Lambda_p$ and $\Lambda_{\pi}$ are chosen to correspond 
to the measured charge radii of the proton and charged pion, which can be found
in Refs.\cite{12,13}. All numerical constants not given explicitly (e.g. the
anomalous magnetic moment $\kappa_p$ of the proton) are taken from Ref.\cite{14}.
The radial RSEs (\ref{eq:14}) are integrated outwards from the origin, with
$V^{pc}$ replaced by $V^{ext}$ or $V^{rel}$ and the factor $f_c$ included.

Even though dipole form factors are used in calculating $\sigma_{0,1}^{ext}$, we 
found that it is quite sufficient for reproducing these phase shifts and
calculating the electromagnetic corrections to use the simple potential for
gaussian charge distributions
\begin{equation}
V^{ext}(r)=\alpha /r \{ \textnormal{erf}(r/c)-1 \} , \label{eq:24}
\end{equation}
where
\[
c^2=2/3 \{ <r^2>_p+<r^2>_{\pi} \}  .
\]
Since $\sigma^{ext}_l$ in Eq.(\ref{eq:21}) contains $m_cf_c=(W^2-m_p^2-\mu_c^2)
/2W$, it is the potential term $2m_cf_cV^{ext}$, with $V^{ext}$ given by
Eq.(\ref{eq:24}), that reproduces $\sigma^{ext}_l$ satisfactorily. 

\begin{figure}
\begin{center}
\includegraphics[height=0.55\textheight,angle=0]{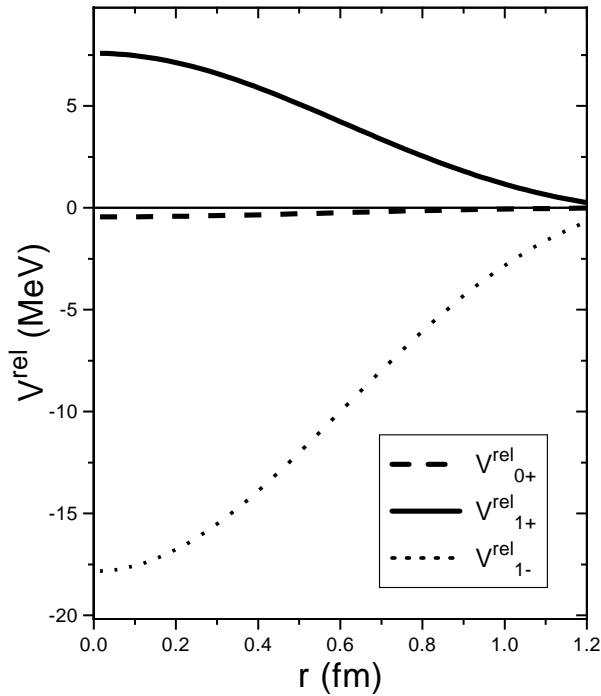}
\caption{The potentials $V_{0+}^{rel}$ and  $V_{1\pm}^{rel}$.}
\label{fig:1}
\end{center}
\end{figure}

For $\sigma^{rel}_{l\pm}$ the numerator $(W^2-m_p^2-\mu_c^2)$ of $m_cf_c$
does not appear explicitly in Eq.(\ref{eq:22}). The potential term can be
written as $2m_cV^{rel}_{l\pm}$ or as $2m_cf_cV^{rel}_{l\pm}$, with
$V^{rel}_{l\pm}$ energy independent in each case, and the phase shifts
$\sigma^{rel}_{0+}$ and $\sigma^{rel}_{1\pm}$ satisfactorily fitted. We checked
that this makes an insignificant difference to the contribution of $V^{rel}_
{l\pm}$ to the electromagnetic corrections. For convenience we therefore took
the term in the form  $2m_cf_cV^{rel}_{l\pm}$. The phase shift
$\sigma^{rel}_{0+}$ is proportional to $q_c$ near threshold and can be
reproduced very well for $T_{\pi} < 100$ MeV by means of a very short range
energy independent potential $V_{0+}^{rel}$. We constructed $V_{1+}^{rel}$ and
$V_{1-}^{rel}$ by adding together a very short range potential and a potential
which has the $r^{-3}$ behaviour for large $r$ characteristic of a spin-orbit
potential and is modified for small $r$ to take account of the charge
distributions. These three potentials are plotted in Fig. \ref{fig:1}.

If we now construct the total electromagnetic potentials 
\begin{equation}
V^{em}_{l\pm}=V^{pc}+V^{ext}+V^{rel}_{l\pm}+V^{vp} \label{eq:25} 
\end{equation}
and insert them in the radial RSEs of the form given in Eq.(\ref{eq:14}), with
$V^{pc}$ replaced by $V^{em}_{l\pm}$, the wavefunction regular at $r=0$ has
the asymptotic behaviour as $r\rightarrow\infty$ 
\[
\sin(q_cr-\eta f_c\ln (2q_cr)-l\pi/2+\sigma_l+\sigma_l^{ext}+\sigma_{l\pm}^
{rel}+\sigma_l^{vp}) .
\]
Here we have neglected higher order contributions in $\alpha$ to the total
electromagnetic phase shifts. The distances at which the asymptotic behaviour
is reached are determined by $V^{vp}$. Since the potentials 
$V^{rel}_{1\pm}$ have an $r^{-3}$ tail one might think that they require 
integration of the appropriate RSEs to the largest distance. It turns out
however that one can stop the integration for 
$V^{rel}_{1\pm}$ at a smaller distance than for $V^{vp}$. 

We now return to Eqs.(\ref{eq:15})-(\ref{eq:17}) and write 
\begin{equation}
\delta_{l\pm}=\sigma_l^{ext}+\sigma_{l\pm}^{rel}+\sigma_l^{vp}+\delta_
{l\pm}^n .
\label{eq:26}
\end{equation}
 The residual phase shifts $\delta_{l\pm}^n$, which arise because of the
hadronic interaction, are usually called nuclear phase shifts, in contrast
to the hadronic phase shifts to be defined later. We now use the identity 
\begin{equation}
e^{2i(\sigma_l-\sigma_0)}(e^{2i\delta_{l\pm}}-1)=e^{2i(\sigma_l-\sigma_0)}
\{e^{2i(\sigma_l^{ext}+\sigma_{l\pm}^{rel}+\sigma_l^{vp})}-1\}+e^{2i\Sigma_
{l\pm}}
(e^{2i\delta_{l\pm}^n}-1) ,  \label{eq:27}
\end{equation}
where
\begin{equation}
\Sigma_{l\pm}=(\sigma_l-\sigma_0)+\sigma_l^{ext}+\sigma_{l\pm}^{rel}+\sigma_l
^{vp} \label{eq:28}
\end{equation}
and
\begin{equation}
\sigma_l-\sigma_0=\sum_{n=1}^{l} \arctan (\frac{\eta f_c}{n}) , \label{eq:29}
\end{equation}
using Eq.(\ref{eq:13}). 
The separation in Eq.(\ref{eq:27}) is made because the phase shifts
$\sigma_{l\pm}^{rel}$ and $\sigma_l^{vp}$ drop off slowly as $l$ increases
and all partial waves must be taken into account. It is convenient to treat
$\sigma_l^{ext}$ in the same way as $\sigma_{l\pm}^{rel}$ 
and  $\sigma_l^{vp}$ and to take the sums indicated in Eqs.(\ref{eq:15}) and
(\ref{eq:16}) over all values of $l$. 
If we work consistently to the lowest order in $\alpha$ ( $\alpha$ for $ext$
and $rel$, $\alpha \eta$ for $vp$), we can replace  
\[
e^{2i(\sigma_l-\sigma_0)}(e^{2i(\sigma_l^{ext}+\sigma_{l\pm}^{rel}+\sigma_l
^{vp})}-1  )
\]
by
\[
(e^{2i\sigma_l^{ext}}-1)+(e^{2i\sigma_{l\pm}^{rel}}-1)+(e^{2i\sigma_l
^{vp}}-1) .
\]
The sum over all partial waves then yields
\begin{equation}
f=f^{pc}+f^{ext}_{1\gamma E}+f^{rel}_{1\gamma E}+f^{vp}+ \sum_{l=0}^{\infty}
\{ (l+1)e^{2i\Sigma_{l+}}f_{l+}^n+le^{2i\Sigma_{l-}}f_{l-}^n \} P_{l},
\label{eq:30}
\end{equation}
\begin{equation}
g=g^{rel}_{1\gamma E}+i \sum_{l=1}^{\infty} ( e^{2i\Sigma_{l+}}f_{l+}^n-
e^{2i\Sigma_{l-}}f_{l-}^n ) P_{l}^{1},   \label{eq:31}
\end{equation}
with
\begin{equation}
f_{l\pm}^n=\frac{e^{2i\delta_{l\pm}^n}-1}{2iq_c} .  \label{eq:32}
\end{equation}
The electromagnetic amplitudes are defined in Eqs.(\ref{eq:7})-(\ref{eq:9}),
(\ref{eq:18}) and (\ref{eq:20}). The electromagnetic phase shifts  
$\Sigma_{l\pm}$ are given in Eqs.(\ref{eq:28}), (\ref{eq:29}) and
(\ref{eq:21})-(\ref{eq:23}).

We want to emphasise that the expressions given in Eqs.(\ref{eq:30}) and
(\ref{eq:31}) are completely sufficient for the phase shift analysis of the
present experimental data for $T_{\pi}\le 100$ MeV. The lowest 
energy at which we shall give results is $T_{\pi}=10$ MeV, for which $\eta
f_c=0.0203$. Present experiments do not go to an energy as low as this, and
the measurements are at angles sufficiently away from the forward direction
that the lowest order approximation to $f^{vp}$ given in Eq.(20) is adequate.

For small angles and energies, the leading correction to the lowest order
approximation to $f^{vp}$ is the imaginary part of the quantity $a^{(1)}_{vp}/
q_{c}$, where $a^{(1)}_{vp}$ is given in Eq. (18.2) of Ref. \cite{11}.
The magnitude of the ratio of this correction to the lowest order approximation
increases as the scattering angle decreases. For $T_{\pi}$ = 10, 15 and 20
MeV the ratio reaches 10\% at values of $\theta$ very close to $3^{\circ}$,
$2^{\circ}$ and $1^{\circ}$ respectively. At $T_{\pi}=15$ MeV, where $\pi^{+}p$
and $\pi^{-}p$ elastic scattering are currently being measured, the magnitude
of this ratio is 6.6\% at $\theta = 5^{\circ}$ and it might just be necessary
to take the correction into account at such a small angle.  

The final step is to introduce the hadronic interaction via effective hadronic
potentials $V_{l\pm}^h$ (in the presence of the electromagnetic interaction)
which are added to $V_{l\pm}^{em}$ :
\begin{equation}
V_{l\pm}=V_{l\pm}^{em}+V_{l\pm}^h  .  \label{eq:33}
\end{equation}
Eq.(\ref{eq:33}) implies that the hadronic potential term has the form
$2m_cf_cV^{h}_{l\pm}$. With $V^{h}_{l\pm}$ chosen to be energy independent,
this means that the term has the specific energy dependence of the factor
$f_c$. It is equally possible to take the hadronic potential term as
$2m_cV^{h}_{l\pm}$, with of course a different energy independent 
potential. For the $\pi^+p$ case we checked that the values of the
electromagnetic corrections are practically independent of the energy
dependence of the hadronic potential term that is chosen. As we shall see in
the next paper, this is not the case for the two-channel $\pi^-p$ situation;
the choice of the energy dependence of the hadronic potential term has
significant consequences for the values of the electromagnetic corrections.
We shall defer to that paper a full
discussion of this delicate point of principle; for the present it is
sufficient to note that, for the $\pi^+p$ corrections, what energy dependence
is chosen for the hadronic potential term is not important. In the
calculation of the corrections given in Section 4 we use RSEs with the form 
\begin{equation}
(\,\frac{d^2}{dr^2}-\frac{l(l+1)}{r^2}+q_c^2-2m_cf_{c}V_{l\pm}(r)\,)\,u_{l\pm}
(r)=0 ,  \label{eq:34}
\end{equation}
with $V_{l\pm}$ given by Eqs.(\ref{eq:33}) and (\ref{eq:25}).

The solutions of the RSEs (\ref{eq:34}) regular at $r=0$ have the asymptotic
behaviour as $r\rightarrow\infty$ 
\[
\sin(q_cr-\eta f_c\ln (2q_cr)-l\pi/2+\sigma_l+\sigma_l^{ext}+\sigma_{l\pm}
^{rel}+\sigma_l^{vp}+\delta_{l\pm}^n) .
\]
The hadronic phase shifts $\delta^h_{l\pm}$ are defined by the asymptotic
behaviour as $r\rightarrow\infty$
\[
\sin{( q_cr-l\pi/2+\delta^h_{l\pm} )}
\]
of the solutions regular at $r=0$ of the RSEs in which $V_{l\pm}$ in
Eq.(\ref{eq:34}) is replaced by $V_{l\pm}^h$. The electromagnetic corrections
$C_{l\pm}$ are then
\begin{equation}
C_{l\pm}=\delta_{l\pm}^n-\delta_{l\pm}^h .  \label{eq:35}
\end{equation}
They arise from the interplay of the potentials $V_{l\pm}^{h}$ and $V_{l\pm}
^{em}$. The quantity $C_{l\pm}$ can, again neglecting terms of higher order
in $\alpha$, be split into four parts:
\begin{equation}
C_{l\pm}=C_{l\pm}^{pc}+C_{l\pm}^{ext}+C_{l\pm}^{rel}+C_{l\pm}^{vp} ,
\label{eq:36}
\end{equation}
where each of the parts arises from the interplay of $V_{l\pm}^{h}$ and the 
corresponding part of $V_{l\pm}^{em}$. We emphasize again that the $V^h$, as
well as $V^{ext}$ and $V^{rel}$, are needed only for the calculation of the
electromagnetic corrections. Once these corrections are determined, the phase
shift analysis, using Eqs.(\ref{eq:30})-(\ref{eq:32}) and (\ref{eq:35}),
determines hadronic phase shifts without any further reference to potentials.
We proceed in the next section to describe the construction of the hadronic
potentials and the evaluation of the corrections. 

\section{Method of evaluation of the corrections}

\begin{figure}
\begin{center}
\includegraphics[height=0.55\textheight,angle=0]{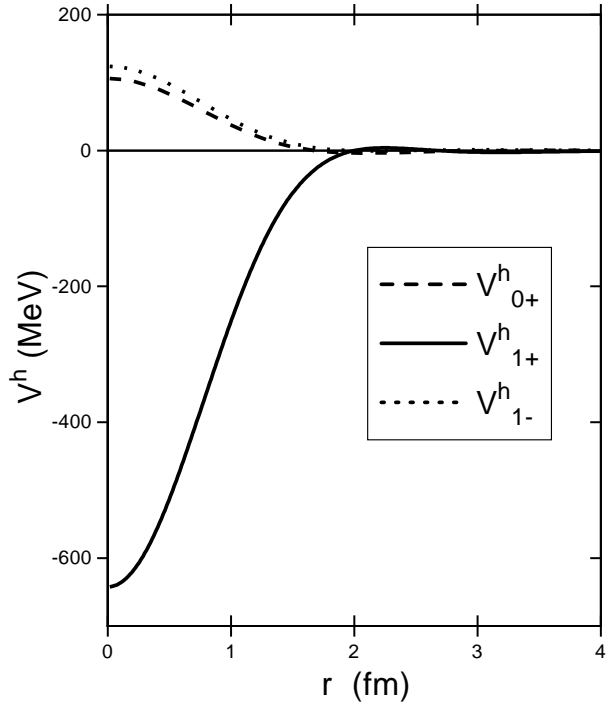}
\caption{The hadronic potentials $V_{0+}^h$ and  $V_{1\pm}^h$.}
\label{fig:2}
\end{center}
\end{figure}

Numerically we evaluated the electromagnetic corrections $C_{l\pm}$ only for
the three lowest partial waves ($0+,\, 1\pm$). This was done in an iterative
procedure of which each step consisted of three stages:\\
a) from values of the hadronic phase shifts $\delta^h_{l\pm}$ we determined
hadronic potentials $V^h_{l\pm}$ which reproduced these $\delta^h_{l\pm}$;\\
b) using these $V^h_{l\pm}$ we calculated values for $C_{l\pm}$;\\
c) with these values of $C_{l\pm}$ we determined via a phase-shift analysis
(PSA) new values for the $\delta^h_{l\pm}$.\\
With these new values of $\delta^h_{l\pm}$ we started again at stage a). The
starting phase shifts for the whole procedure are of no great importance;
in practice we used the output from the analysis of Arndt {\it et al}.
\cite{15}. The iterative procedure was continued until the $C_{l\pm}$, and
therefore the $\delta^h_{l\pm}$, achieved stable values; three iteration steps
were found to be sufficient. We now describe the three stages in more detail. 

a) For each  $V^h_{l\pm}$ we used a parametric form containing a range
parameter which was fixed at 1 fm. This form contains three further parameters 
for each partial wave. For a fixed set of these parameters we inserted
$V^h_{l\pm}$ instead of $V_{l\pm}$ in Eq.(\ref{eq:34}) and by 
numerical integration obtained the solutions $u^h_{l\pm}(r)$ regular at $r=0$.
Outside the range of $V^h_{l\pm}$ the asymptotic form as $r\rightarrow\infty$
of the  $u^h_{l\pm}(r)$,
\begin{equation}
 u^h_{l\pm}(r) \sim   a^h_{l\pm}j_l(q_cr)+b^h_{l\pm}n_l(q_cr) ,\label{eq:37}
\end{equation}
gave the values of $\delta^h_{l\pm}$ generated by the $V^h_{l\pm}$ via the
equation 
\begin{equation}
 \tan{\delta^h_{l\pm}}=\frac{b^h_{l\pm}}{a^h_{l\pm}} .\label{eq:38}
\end{equation}
The three parameters in each $V^h_{l\pm}$ were then varied in order to get
the best possible fit to the values of $\delta^h_{l\pm}$ up to $T_{\pi}=100$
MeV. The potentials constructed in this way reproduced these values within 
their experimental errors. In Fig. \ref{fig:2} we show the potentials
$V^h_{0+}$ and $V^h_{1\pm}$ for the final step of the iteration. 
The analytical form and further details can be found in Ref.\cite{16}.

b) Using the $V^h_{l\pm}$ constructed in a) the electromagnetic corrections
were evaluated in the following way. 
We integrated Eq.(\ref{eq:34}) with the full potentials $V_{l\pm}$ and
obtained the solutions $u_{l\pm}$ regular at $r=0$, integrating outwards to
a distance where $V^h_{l\pm}+V^{ext}+V^{rel}_{l\pm}+V^{vp}$ is negligible 
compared to $V^{pc}$. In this region the asymptotic form of the $u_{l\pm}$,
\begin{equation}
 u_{l\pm}(r) \sim   a_{l\pm}F_l(\eta f_c;q_cr)+b_{l\pm}G_l(\eta f_c;q_cr) ,
\label{eq:39}
\end{equation}
where $F_l$ and $G_l$ are the standard point charge Coulomb wavefunctions,
yielded the nuclear phase shifts $\delta^n_{l\pm}$ via the equation
\begin{equation}
 \tan{(\sigma_l^{ext}+\sigma^{rel}_{l\pm}+\sigma_l^{vp}+\delta^n_{l\pm}) }
=\frac{b_{l\pm}}{a_{l\pm}} .
\label{eq:40}
\end{equation}
Eq.(\ref{eq:35}) then gave the corrections $C_{l\pm}$ for this step of the
iteration procedure, the $\delta^h_{l\pm}$ being of course the hadronic phase
shifts generated by the potentials $V^h_{l\pm}$. Note that the use of the
asymptotic forms (\ref{eq:37}) and (\ref{eq:39}) instead of those given just
before Eq.(\ref{eq:35}) enables the numerical integration to go to smaller
distances.

Though the PSA of the $\pi^+ p$ data is not the topic of this paper we see
that because of the iterative procedure we have described, the PSA goes hand
in hand with the evaluation of the electromagnetic corrections. We therefore
describe briefly how the PSA is done; a full description will be given in a
separate paper.

c) The PSA uses Eqs.(\ref{eq:30}) and (\ref{eq:31}) with the sums over $l$
taken to $l=3$. For $l=2,3$ the small and poorly known nuclear phase shifts
$\delta^n_{l\pm}$ need to be taken from a PSA at higher energies and then
extrapolated to the lower energies with which we are concerned. Taking the
electromagnetic corrections $C_{l\pm}$ as calculated in stage b), 
the PSA determined the three hadronic phase shifts $\delta^h_{0+}$,
$\delta^h_{1\pm}$. Parametric forms for these phase shifts were used and the
parameters varied to obtain the best fit to the data. The parametric forms are 
simple low energy expansions in the pion c.m. kinetic energy. Three parameters
were used for the $s$-wave  and two for the non-resonant $p$-wave. For the
resonant phase shift $\delta^{h}_{1+}$, a parametrised background term was
added to a Breit-Wigner resonant term with an energy dependent width. The
hadronic potentials constructed in a) can be considered as an auxiliary
alternative parametrisation of the hadronic phase shifts which is used only
for the calculation of the electromagnetic corrections.

A very important observation of a conceptual nature needs to be made.
We have called the quantities $\delta_{l\pm}^h$ hadronic phase shifts, but
they are obtained from a phase shift analysis which contains the \textit
{physical} masses of $\pi^+$ and $p$. It is therefore clear that the
$\delta_{l\pm}^h$ cannot be considered to be strictly hadronic quantities. 
Such quantities relate to a situation in which the electromagnetic interaction
is switched off ($\alpha =0$) and therefore need to be obtained using the
hadronic masses of $\pi^+$ and $p$. These hadronic masses are not the same
as their physical masses (indeed it is universally accepted that the hadronic
mass of $\pi^+$ is very close to the physical mass of $\pi^0$). In the same
sense we have noted that the quantities $V_{l\pm}^h$ are effective hadronic
potentials in the presence of the electromagnetic interaction. They would be
different in the complete absence of all electromagnetic interactions.
Therefore in our present work, as in all previous work known to us, we avoid
speculation about the strictly hadronic situation and give quantities that
we call hadronic phase shifts and electromagnetic corrections which, though
they have a precise definition within the framework of our potential model,
are not the full corrections that would give truly hadronic phase shifts 
by subtraction from the nuclear phase shifts. Any attempt to completely purge
the experimental data of all electromagnetic effects would be very
speculative. Our aim is the more modest one of taking account of those
electromagnetic effects that may be calculated with reasonable confidence.

\section{Numerical results for the corrections}

The final results for the three electromagnetic corrections $C_{0+}$,
$C_{1\pm}$ are given in Table \ref{tab:1} from 10 to 100 MeV pion lab
kinetic energy. Estimates of the uncertainties in the values of these
corrections are also given in the table. In making these estimates it is 
necessary to distinguish between a) uncertainties
coming from applying our model and b) uncertainties coming from the choice of
the model itself.\
\begin{table}
\begin{center}
\caption{Values in degrees of the electromagnetic corrections $C_{0+}$,
$C_{1-}$ and $C_{1+}$ 
as functions of the pion lab kinetic energy $T_{\pi}$ (in MeV).}
\label{tab:1}
\begin{tabular}{|c|c|c|c|}
\hline
 $T_{\pi}$ & $C_{0+}$ & $C_{1+}$ &$C_{1-}$ \\\hline
10  & 0.081$\pm$ 0.003&-0.023$\pm$ 0.000&0.005$\pm$ 0.000\\
15  & 0.083$\pm$ 0.003&-0.034$\pm$ 0.000&0.007$\pm$ 0.000\\
20  & 0.085$\pm$ 0.004&-0.047$\pm$ 0.001&0.009$\pm$ 0.000\\
25  & 0.088$\pm$ 0.005&-0.060$\pm$ 0.001&0.010$\pm$ 0.000\\
30  & 0.090$\pm$ 0.005&-0.074$\pm$ 0.002&0.012$\pm$ 0.000\\
35  & 0.093$\pm$ 0.006&-0.090$\pm$ 0.002&0.013$\pm$ 0.001\\
40  & 0.096$\pm$ 0.006&-0.106$\pm$ 0.003&0.014$\pm$ 0.001\\
45  & 0.099$\pm$ 0.007&-0.125$\pm$ 0.003&0.015$\pm$ 0.001\\
50  & 0.101$\pm$ 0.007&-0.145$\pm$ 0.004&0.016$\pm$ 0.001\\
55  & 0.104$\pm$ 0.008&-0.168$\pm$ 0.005&0.017$\pm$ 0.002\\
60  & 0.107$\pm$ 0.009&-0.194$\pm$ 0.006&0.018$\pm$ 0.002\\
65  & 0.109$\pm$ 0.010&-0.223$\pm$ 0.008&0.018$\pm$ 0.002\\
70  & 0.111$\pm$ 0.010&-0.255$\pm$ 0.011&0.019$\pm$ 0.002\\
75  & 0.114$\pm$ 0.011&-0.291$\pm$ 0.016 &0.019$\pm$ 0.003\\
80  & 0.115$\pm$ 0.012&-0.332$\pm$ 0.023 &0.019$\pm$ 0.003\\
85  & 0.118$\pm$ 0.014&-0.378$\pm$ 0.032 &0.019$\pm$ 0.004\\
90  & 0.119$\pm$ 0.015&-0.429$\pm$ 0.045 &0.019$\pm$ 0.004\\
95  & 0.121$\pm$ 0.016&-0.485$\pm$ 0.062 &0.019$\pm$ 0.004\\
100 & 0.123$\pm$ 0.018&-0.547$\pm$ 0.088 &0.020$\pm$ 0.005\\ \hline
\end{tabular}
\end{center}
\end{table}

One source of a) is the uncertainty in the hadronic phase shifts, which comes
from the PSA of the experimental data (experimental errors in the data and
uncertainty in the $d$- and $f$-wave phase shifts used as input). The
uncertainties in the hadronic phase shifts will be given in full in the
separate paper on the PSA that was mentioned earlier. The resulting
uncertainties in the corrections were thoroughly studied by keeping track of
the corrections obtained using a variety of input values for the hadronic
phase shifts (in particular for the successive iteration steps described in
Section 3). These uncertainties are substantially smaller than those coming
from the use of the particular form of the parametrised hadronic potentials,
which are also of type a). For the final numerical calculations we fixed the 
range parameter in the potentials at 1 fm. Its value can be varied from 0.8
fm to 1.2 fm without any significant change in the quality of the fits to the
hadronic phase shifts. In this way we obtained an estimate of the uncertainties
coming from the fact that some of the fine details of the potentials may have
been missed.

The estimated uncertainties given in Table \ref{tab:1} were obtained by
combining the two sources of type a) described in the previous paragraph. For
$C_{0+}$ and $C_{1-}$ these uncertainties are much smaller than the errors on
the hadronic phase shifts that arise from the PSA itself. For example, at 100
MeV the errors on $\delta_{0+}$ and $\delta_{1-}$  are $0.15^{\circ}$ and
$0.09^{\circ}$, compared with the uncertainties in the corrections of
$0.018^{\circ}$ and $0.005^{\circ}$ respectively. The situation is different
for $C_{1+}$, where the error
on $\delta_{1+}$ at 100 MeV is $0.05^{\circ}$, while the correction is
$-0.547^{\circ}$ and its uncertainty is $0.088^{\circ}$. The correction itself
is very large ($\delta_{1+}$ itself is $21.27^{\circ}$ at this energy) and the
uncertainty in the correction is somewhat larger than the error arising from
the PSA.

The corrections in Table \ref{tab:1} are intended for use in future PSAs of
experimental data. Such PSAs need to use Eqs (\ref{eq:30})-(\ref{eq:32}),
(\ref{eq:28}) and (\ref{eq:35}) as well as higher order corrections to $f^{vp}$
and $\sigma^{vp}$  for experiments  at very low energies and angles, as
discussed in Section 2. For the output phase shift $\delta_{1+}$ the
uncertainty in the correction $C_{1+}$ needs to be taken into account as well
as the statistical error arising from the PSA itself.

Uncertainties of type b) arise from the choice of the RSEs (\ref{eq:34}),
with the hadronic potential term having the specific energy dependence of the
factor $f_c$. We have already remarked that the choice of the energy
dependence of this term has no significant effect on the corrections.
The RSEs themselves are the only relativistic equations that are well suited
to a two-body problem. It is not meaningful to give an uncertainty arising from
this source; all one can do is to compare the results obtained using our
present model with those from alternative models, as we do in a moment.

\begin{table}
\begin{center}
\caption{Values in degrees of the various contributions to the electromagnetic
corrections $C_{0+}$ as functions of the pion lab kinetic energy $T_{\pi}$
(in MeV).}
\label{tab:2}
\begin{tabular}{|c|c|c|c|c|c|}
\hline
$T_{\pi}$ & $C_{0+}^{pc}$ & $C_{0+}^{ext}$ & $C_{0+}^{rel}$ & $C_{0+}^{vp}$\\
\hline
 10 &  0.093 & -0.008 & -0.002 & 0.000 \\
 20 &  0.102 & -0.013 & -0.004 & 0.000 \\
 30 &  0.113 & -0.018 & -0.005 & 0.000 \\
 40 &  0.123 & -0.022 & -0.005 & 0.000 \\
 50 &  0.134 & -0.027 & -0.006 & 0.000 \\
 60 &  0.145 & -0.031 & -0.007 & 0.000 \\
 70 &  0.155 & -0.036 & -0.008 & 0.000 \\
 80 &  0.165 & -0.041 & -0.009 & 0.000 \\
 90 &  0.175 & -0.046 & -0.010 & 0.000 \\
 100 &  0.183 & -0.051 & -0.010 & 0.000 \\ \hline

\end{tabular}
\end{center}
\end{table}

\begin{table}
\begin{center}
\caption{Values in degrees of the various contributions to the electromagnetic
corrections $C_{1-}$ and $C_{1+}$ as functions of the pion lab kinetic energy
$T_{\pi}$ (in MeV).}
\label{tab:3}
\begin{tabular}{|c|c|c|c|c|c|}
\hline

$T_{\pi}$ & $C_{1+}^{pc}$ & $C_{1+}^{ext}$ & $C_{1+}^{rel}$ & $C_{1+}^{vp}$ \\
\hline
  10 & -0.023 &  0.000 &  0.000 & -0.000 \\
 20 & -0.047 &  0.001 & -0.000 & -0.000 \\
  30 & -0.074 &  0.001 & -0.001 & -0.001 \\
 40 & -0.104 &  0.002 & -0.004 & -0.001 \\
  50 & -0.141 &  0.004 & -0.008 & -0.001 \\
  60 & -0.184 &  0.007 & -0.016 & -0.001 \\
 70 & -0.236 &  0.011 & -0.029 & -0.002 \\
  80 & -0.300 &  0.017 & -0.048 & -0.002 \\
  90 & -0.376 &  0.025 & -0.077 & -0.003 \\
 100 & -0.464 &  0.036 & -0.119 & -0.003 \\ \hline
$T_{\pi}$ & $C_{1-}^{pc}$ & $C_{1-}^{ext}$ & $C_{1-}^{rel}$ & $C_{1-}^{vp}$ \\
\hline
 10 &  0.004 & -0.000 &  0.000 & 0.000 \\
 20 &  0.008 & -0.000 &  0.000 & 0.000 \\
 30 &  0.011 & -0.000 &  0.000 & 0.000 \\
 40 &  0.014 & -0.000 &  0.000 & 0.000 \\
 50 &  0.016 & -0.000 &  0.000 & 0.000 \\
 60 &  0.018 & -0.000 & -0.000 & 0.000 \\
 70 &  0.019 & -0.000 & -0.001 & 0.000 \\
 80 &  0.021 & -0.001 & -0.001 & 0.000 \\
 90 &  0.022 & -0.001 & -0.002 & 0.000 \\
100 &  0.024 & -0.001 & -0.004 & 0.000 \\ \hline
\end{tabular}
\end{center}
\end{table}

In Tables \ref{tab:2} and \ref{tab:3} we give (at a smaller set of energies)
the various contributions to the corrections (Eq.(\ref{eq:11})).These
components are additive to a very good approximation and tiny differences
between the sums of the numbers in these tables and the complete corrections
in Table \ref{tab:1} are due to higher order effects. The corrections are
clearly dominated by the component $C^{pc}$ arising from the interplay between
the hadronic and point charge Coulomb potentials. The only other significant
components are $C_{0+}^{ext}$ and $C_{1+}^{rel}$ near $T_{\pi}=100$ MeV.

\begin{figure}
\begin{center}
\includegraphics[height=0.55\textheight,angle=0]{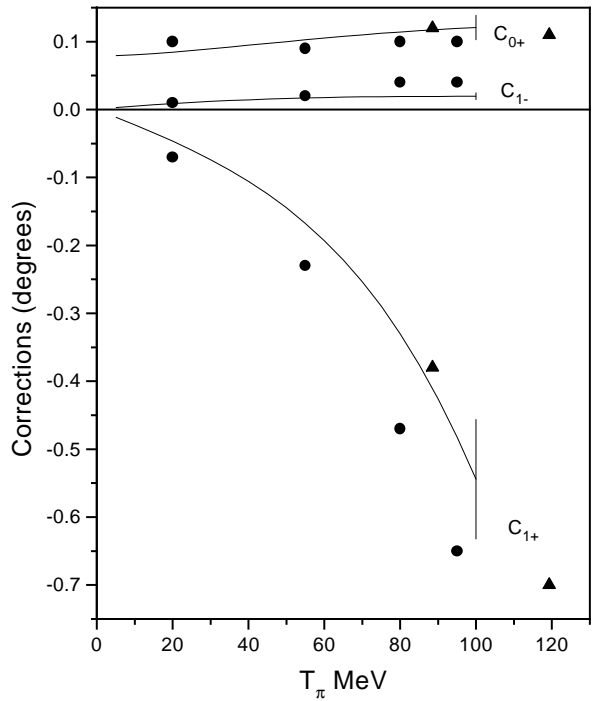}
\caption{Values in degrees of the electromagnetic corrections $C_{0+}$,
$C_{1+}$ and $C_{1-}$ from our present calculation (solid curves), from
NORDITA \cite{3} (circles) and Zimmermann \cite{8} (triangles).}
\label{fig:3}
\end{center}
\end{figure}

Our results are compared with those of the dispersion theory approach used by
NORDITA\cite{3} in Fig.\ref{fig:3}. They have not included the vacuum
polarisation contribution but we see from Tables \ref{tab:2} and \ref{tab:3}
that it is negligible for each of the partial waves at the energies we
consider. The results of Zimmermann \cite{8}, who uses a slightly different
version of the potential model, are also shown in Fig. \ref{fig:3}. We have
added our values of the relativistic corrections to the results of Zimmermann.
We have indicated in Fig.\ref{fig:3} the uncertainties in our corrections at
$100$ MeV, as given in Table \ref{tab:1}. No errors are given in Refs. \cite{3}
 and \cite{8} for the corrections presented there.
For $C_{0+}$  the results are in very good agreement, the differences being 
considerably smaller than the error on $\delta^{h}_{0+}$ arising from the PSA.
For $C_{1+}$ our results agree quite well with those of Zimmermann, which
indicates the stability of the results obtained with the potential model.
However, there is a systematic difference from the results of NORDITA, the
discrepancy increasing with energy. At 100 MeV the difference is about 2.5
times our estimated uncertainty in $C_{1+}$ and over 4 times the error in
$\delta_{1+}$ arising from the PSA. For $C_{1-}$  our results agree with those
of NORDITA at low energies, but at 100 MeV the discrepancy is roughly twice
the error in $\delta^{h}_{1-}$. However, since $\delta^{h}_{1-}$ is very small,
the corrections are of minor importance for the results of the PSA.    

As we discussed in Section 1, the calculation of electromagnetic corrections 
using dispersion relations omits what could be important 
medium range effects due to $t$-and $u$-channel exhanges, which the potential
model includes quite reliably. 
The differences of our results from those of NORDITA are probably due to such
medium range effects. In particular the stronger energy dependence of the
NORDITA results for $C_{1+}$ shown in Fig. \ref{fig:3} is likely to be a 
medium range effect, perhaps due to their omission of
$t$-channel $\pi\pi$ ($T=0$, $J=0$) exchange. We therefore claim a higher
degree of reliability for our present calculation of the electromagnetic
corrections, compared with that of NORDITA. At the same time it is pleasing
that the corrections in the $\pi^+p$ case are very reliably known and are to
a large extent independent of the model used for their calculation.

\begin{ack} 

We thank the Swiss National Foundation and PSI (`Paul Scherrer Institut') for
financial support. We are also indebted to two referees for several very
helpful comments and suggestions. 

\end{ack}

\end{document}